\documentclass[format=acmsmall, review=false, screen=true]{acmart}

\usepackage{booktabs} 

\usepackage{courier}
\usepackage{amsmath}
\usepackage{amssymb}
\usepackage{amsfonts}
\usepackage{xcolor}
\usepackage{multirow}
\usepackage{float}
\usepackage{graphicx}
\usepackage{graphics}
\usepackage{subfigure}
\usepackage{float}
\usepackage{bigstrut}
\usepackage{booktabs}
\usepackage{bibentry}
\usepackage{balance}

\usepackage[ruled]{algorithm2e} 

\SetAlFnt{\small}
\SetAlCapFnt{\small}
\SetAlCapNameFnt{\small}
\SetAlCapHSkip{0pt}
\IncMargin{-\parindent}

\acmJournal{TIST}
\acmVolume{9}
\acmNumber{4}
\acmArticle{39}
\acmYear{2010}
\acmMonth{3}
\copyrightyear{2009}

\setcopyright{acmlicensed}

\acmDOI{0000001.0000001}

\received{February 2007}
\received[revised]{March 2009}
\received[accepted]{June 2009}

\begin{document}
\title[An Attention-based Approach for Identification of Misinformation]{Mining Significant Microblogs for Misinformation Identification: An Attention-based Approach}
\author{Qiang Liu, Feng Yu, Shu Wu, and Liang Wang}
\affiliation{%
  \institution{Chinese Academy of Sciences}}

\begin{abstract}
With the rapid growth of social media, massive misinformation is also spreading widely on social media, such  as microblog, and bring negative effects to human life. Nowadays, automatic misinformation identification has drawn attention from academic and industrial communities. For an event on social media usually consists of multiple microblogs, current methods are mainly based on global statistical features. However, information on social media is full of noisy and outliers, which should be alleviated. Moreover, most of microblogs about an event have little contribution to the identification of misinformation, where useful information can be easily overwhelmed by useless information. Thus, it is important to mine significant microblogs for a reliable misinformation identification method. In this paper, we propose an Attention-based approach for Identification of Misinformation (AIM). Based on the attention mechanism, AIM can select microblogs with largest attention values for misinformation identification. The attention mechanism in AIM contains two parts: content attention and dynamic attention. Content attention is calculated based textual features of each microblog. Dynamic attention is related to the time interval between the posting time of a microblog and the beginning of the event. To evaluate AIM, we conduct a series of experiments on the Weibo dataset and the Twitter dataset, and the experimental results show that the proposed AIM model outperforms the state-of-the-art methods.
\end{abstract}

%
%

\ccsdesc[500]{Information systems~Data mining}
\ccsdesc[500]{Information systems~Collaborative and social computing systems and tools}
\ccsdesc[300]{Information systems~Web mining}
\ccsdesc[300]{Information systems~Social networks}

%
%

\keywords{Misinformation identification, attention model, social media, significant microblogs}

\thanks{Author's addresses: Q. Liu, F. Yu, S. Wu and L. Wang, the Center for Research on Intelligent Perception and Computing (CRIPAC), National Laboratory of Pattern Recognition (NLPR), Institute of Automation, Chinese Academy of Sciences (CASIA) and the University of Chinese Academy of Sciences (UCAS), Beijing 100080, China; emals: \{qiang.liu, feng.yu, shu.wu, wangliang\}@nlpr.ia.ac.cn.}

\maketitle

\section{Introduction}

With the rapid growth of social media, such as Facebook, Twitter, and Weibo, people are sharing information and expressing their attitudes publicly. Social media brings great convenience to users, and information can be spread rapidly and widely nowadays. However, misinformation can also be spread on the Internet more easily. Misinformation brings significant harm to daily life, social harmony, or even public security. With the growth of the Internet and social media, such harm will also grow greater. For instance, as the loss of MH370 has drawn worldwide attention, a great amount of rumors has spread on social media, e.g., MH370 has landed in China \footnote{http://www.fireandreamitchell.com/2014/03/07/rumor-malaysia-airlines-mh370-landed-china/}, the loss of MH370 is caused by terrorists \footnote{http://www.csmonitor.com/World/Asia-Pacific/2014/0310/Malaysia-Airlines-flight-MH370-China-plays-down-terrorism-theories-video}, and Russian jets are related to the loss of MH370 \footnote{http://www.inquisitr.com/1689765/malaysia-airlines-flight-mh370-russian-jets-in-baltic-may-hold-clue-to-how-flight-370-vanished/}. These rumors about MH370 mislead public attitudes to a wrong direction and delay the search of MH370. Up to March 15, 2017, on the biggest Chinese microblog website Sina Weibo \footnote{http://weibo.com}, 32,076 rumors have been reported and collected in its misinformation management center \footnote{http://service.account.weibo.com/?type=5\&status=4}. Accordingly, it is crucial to evaluate information credibility and to detect misinformation on social media.

Nowadays, to automatically identify misinformation on social media, some methods have been recently proposed. Usually, and event, which may be misinformation or true information, contains a group of microblogs. Thus, most of existing methods identify misinformation at the microblog level \cite{castillo2011information,qazvinian2011rumor,gupta2013faking} or the event level \cite{kwon2013prominent,zhao2015enquiring,ma2015detect}. Some studies investigate the aggregation of credibility from the microblog level to the event level \cite{jin2014news}. On the contrary, considering dynamic information, some work designs temporal features based on the prorogation properties over time \cite{kwon2013prominent} or trains a model with features generated from different time periods \cite{ma2015detect}. Recently, Recurrent Neural Networks (RNN) \cite{mikolov2010recurrent} have been incorporated for misinformation identification, and the Gated Recurrent Unit (GRU) structure \cite{chung2015gated} is proved to have satisfactory performance \cite{ma2016detecting}. Moreover, some methods take usage of users' feedbacks (comments and attitudes) to evaluate the credibility \cite{giudice2010crowdsourcing,rieh2014audience,zhao2015enquiring}. For instance, \cite{zhao2015enquiring} takes out signal tweets, which indicates users' skepticism about factual claims for detecting misinformation.

Though above methods succeed in misinformation identification, they have severe drawbacks. Among these methods, some identify misinformation according to global statistical features of an event or a time window, some calculate the credibility of each microblog and then aggregate them to the the credibility of the whole event. However, information on social media is full of noisy and outliers, which should be alleviated. Moreover, most of microblogs about an event have little contribution to the identification of misinformation. As shown in the example of misinformation on Sina Weibo in Table \ref{tab:example}, most users are simply reposting the fake news, or credence to the misinformation. Only a few users express their questions about the misinformation. Thus, it is important to select those significant microblogs, and obtain a reliable misinformation identification method.

\begin{table}[tb]
  \centering\scriptsize
  \caption{An example of misinformation on Sina Weibo.}
  {
    \begin{tabular}{cl}
    \toprule
    posting time & \multicolumn{1}{c}{content} \\
    \midrule
    \multirow{2}[0]{*}{2014/03/20 23:55} & Hearing from an Australian friend: The plane has been found in the international waters \\
          & near Perth. It is proven to be MH370 according to a major component of the plane. \\
    2014/03/01 23:56 & May God bless them! \\
    2014/03/20 23:57 & Reposting \\
    2014/03/20 23:58 & It is serious to spread rumors! \\
    2014/03/20 23:59 & Reposting \\
    2014/03/21 00:00 & Hopefully it's not true. \\
    2014/03/21 00:02 & Reposting \\
    2014/03/21 00:03 & Really??? \\
    2014/03/21 00:04 & Waiting for official confirmation tomorrow. \\
    2014/03/21 00:06 & Reposting \\
    2014/03/21 00:07 & Reposting \\
    2014/03/21 00:17 & Let¡¯s watch the exact news tomorrow morning. Anyway, may God bless them! \\
    2014/03/21 00:18 & Reposting \\
    2014/03/21 00:21 & What a bad news! \\
    2014/03/21 00:22 & Reposting \\
    2014/03/21 00:32 & Reposting \\
    2014/03/21 00:35 & Reposting unreliable information, what an expert! \\
    2014/03/21 00:46 & Reposting \\
    2014/03/21 00:51 & No! No! No! \\
    2014/03/21 01:06 & Dare to post misinformation! \\
    2014/03/21 01:09 & Reposting \\
    2014/03/21 01:25 & Is it reliable? \\
    2014/03/21 01:32 & Reposting \\
    \bottomrule
    \end{tabular}%
  }
  \label{tab:example}%
\end{table}%

Fortunately, the attention mechanism \cite{itti1998model} is suitable for selecting most significant components of information. Via the attention mechanism, components which contribute more to a specific task have large weights for satisfying the objective as much as possible. The attention mechanism has succeed in multiple tasks, such as visual object detection \cite{ba2015multiple}, image caption \cite{xu2015show}, machine translation \cite{bahdanau2014neural}, text summarization \cite{rush2015neural} and text classification \cite{wang2016attention}. Accordingly, with the attention mechanism, we are able to mine signification microblogs for identifying misinformation, and design a reliable automatic detection method.

Moreover, misinformation early detection is another important and practical task, in which we need to detect misinformation as early as possible \cite{zhao2015enquiring,ma2016detecting}. Thus, we can take immediate actions at the beginning stage of spreading of misinformation, and minimize the baneful influence. For early detection, we need to identify misinformation with the first several microblogs. And with the attention mechanism, we can identify misinformation with several significant microblogs. Accordingly, the attention mechanism is naturally suitable for misinformation early detection.

In this work, we propose an Attention-based approach for Identification of Misinformation (\textbf{AIM}). First, for each microblog belonging to an event, we calculate corresponding attention value based on its textual features. This attention value is named as content attention. Second, considering microblogs posted at different time have distinct significance for the event, we calculate dynamic attention for each microblog. Dynamic attention can be determined related to the time interval between the posting time of a microblog and the beginning of the event. Then, we aggregate the content attention and the dynamic attention, and obtain the final attention weights for each microblog belonging to an event. Weighted sum of these microblogs can be performed to generate the representation of the whole event. Finally, the prediction of misinformation or true information can be made based on the event representation.

In summary, the main contributions of this work are listed as follows:

\begin{itemize}
\item
We incorporate the attention mechanism for misinformation identification on social media, which mines the most significant microblogs.

\item
We design both content attention and dynamic attention, for capturing different aspects of significance of microblogs for misinformation identification.

\item
Experiments conducted on two real-world datasets, i.e., the Weibo dataset and the Twitter dataset, show that AIM is effective and outperforms state-of-the-art methods significantly.

\item
Visualization of the leaned attention mechanism in AIM demonstrates the rationality of our proposed method.

\end{itemize}

The rest of this paper is organized as follows. In section 2, we review some related work on misinformation identification and attention mechanism. Then we detail the proposed AIM model in section 3. In section 4, we conduct and analyze experiments on two real-world datasets, and compare with several state-of-the-art methods. In section 5, we illustrate some visualization examples of the leaned attention mechanism. Section 6 concludes this work and discusses future research directions.

\section{Related Work}

In this section, we briefly review some related works on misinformation identification and attention mechanism.

\subsection{Misinformation Identification}

Recently, many methods have been put forward for misinformation automatic identification. The work of \cite{kumar2016disinformation} analyzes impact and characteristics of hoax articles in Wikipedia and proposes an efficient method to identify these Wikipedia hoaxes. On social media, some researchers identify misinformation at the post level \cite{castillo2011information,qazvinian2011rumor}, i.e., classifying a single microblog post as being credible or not based on tweet-based features. Some perform a characterization analysis for the spread of fake images of microblog posts during crisis events \cite{gupta2013faking}. Some identify whether an event belongs to misinformation or truth information and extract handcrafted features from the event level \cite{kwon2013prominent,ma2015detect,zhao2015enquiring,wu2016information}. Another work obtains credibility of a microblog post and then aggregates credibility to the event level \cite{jin2014news}.
Moreover, some other works extract more effective handcrafted features. For instance, the work of \cite{jin2016news} takes advantage of "wisdom of crowds" to identify fake news, i.e., mining opposing voices from conflicting viewpoints. Based on the time series of misinformation lifecycle, the temporal characteristics of social context information are captured in \cite{kwon2013prominent,ma2015detect}. The work of \cite{giudice2010crowdsourcing,rieh2014audience} investigate the web page credibility through users' feedback. Signals tweets are identified from trending misinformation via finding signature text phrases expressing skepticism about factual claims \cite{zhao2015enquiring}. Recently, a RNN based model attempts to capture the dynamic temporal signals in the misinformation diffusion process and incrementally learn both the temporal and textual representations of an event \cite{ma2016detecting}.

\subsection{Attention Mechanism}

Attention mechanism is first applied to a visual attention system for rapid scene analysis \cite{itti1998model}. The visual attention system selects attended locations in order of decreasing saliency, so that a complex scene can be understood by rapidly selecting saliency locations in a computationally efficient method. In recent years, Deep Neuron Network (DNN) is getting increasingly popular. Attention mechanism is once again taken out to be integrated into Deep Neural Networks (DNN). Attention mechanism is incorporated into RNN in \cite{mnih2014recurrent}, to attend to different locations within the images one at a time and process them sequentially. The attention mechanism can help control expensive computation independent of the input image size and learn tracking without explicit training signals.

Furthermore, the work of \cite{ba2015multiple}  extends the attention RNN model to multiple objects detection task, that is learning to both localize and recognize multiple objects despite being given only class labels. For an image caption task, a attention-based model is able to automatically fix its attention on salient objects of an input image while generating the corresponding words of the output sentence \cite{xu2015show}. Some employ attention mechanism in a visual question answering task, such as generating question-guided attention to image feature maps for each question \cite{chen2015abc}, a question-guided spatial attention to images for questions of spatial inference \cite{xu2016ask} and querying an image and inferring the answer multiple times to narrow down the attention to images progressively via stacked attention networks \cite{yang2016stacked}. For fine-grained image classification, an attention-based Convolutional Neural Networks (CNN) model improves the performance of which to attend and what to extract without expensive annotations like bounding box or part information \cite{xiao2015application}.

In the field of natural language processing, researchers first introduce attention mechanism to neural machine translation. Based on a primitive encoder-decoder architecture, the work of \cite{bahdanau2014neural} searches a source sentence to attend to the most relevant words to predict a target word. Some extend the attention mechanism to global and local ones and compare different methods of obtaining attention scores \cite{luong2015effective}. Moreover, a hierarchical attention mechanism guides layers in a CNN model to models text in \cite{yin2016attention}. The work of \cite{dhingra2016gated} integrates a multi-hop architecture with a gated-attention layer based on multiplicative interactions between the query embedding and the intermediate states of a recurrent document reader. Besides, attention mechanism is introduced into more research issues, such as abstractive text summarization \cite{rush2015neural}, text comprehension task \cite{dhingra2016gated,kadlec2016text,yin2016attention}, relation classification \cite{wang2016relation,zhou2016attention} and text classification \cite{yang2016hierarchical}. In \cite{chorowski2015attention}, a novel model for speech recognition is proposed, which incorporates both content-based attention \cite{bahdanau2014neural,xu2015show} and location-based attention \cite{graves2013generating}.

\section{Proposed Method}

In this section, we first formulate the problem. Then, we detail the proposed AIM model. Finally, we present the parameter learning procedure for the AIM model.

\subsection{Problem Formulation}
The problem to be studied in this paper can be formulated as follows. Suppose a set of events are denoted as $E = \left\{ {e_1 ,e_2 ,...,e_n } \right\}$. $l_{e_i}$ is the label of the corresponding event $e_i$, where $l_{e_i} = 1$ means event $e_i $ is misinformation and $l_{e_i }  = 0$ otherwise. The microblogs of the event $e_i$ can be denoted as $M^{e_i }  = \left[ {m_1^{e_i } ,m_2^{e_i } ,...,m_{n_{e_i}}^{e_i }} \right]$, where $n_{e_i}$ is the number of microblogs of this event. All microblog sets can be written as $M = \left\{ {M^{e_1 } ,M^{e_2 } ,...,M^{e_n } } \right\}$.
Each microblog $m_j^{e_i }$ consists of its content feature vector $\mathbf{f}_j^{e_i}$ and posting time $t_j^{e_i}$. Then, content feature vectors and posting time of event $e_i $ can be denoted as $\mathbf{F}^{e_i }  = \left[ {\mathbf{f}_1^{e_i } ,\mathbf{f}_2^{e_i } ,...,\mathbf{f}_{n_{e_i}}^{e_i }} \right]$ and $T^{e_i }  = \left[ {t_1^{e_i } ,t_2^{e_i } ,...,t_{n_{e_i}}^{e_i }} \right]$ respectively. In this work, our task is to identify whether an event on social media is misinformation or not.

\begin{figure}[!tb]
\centering
\includegraphics[width=1\linewidth]{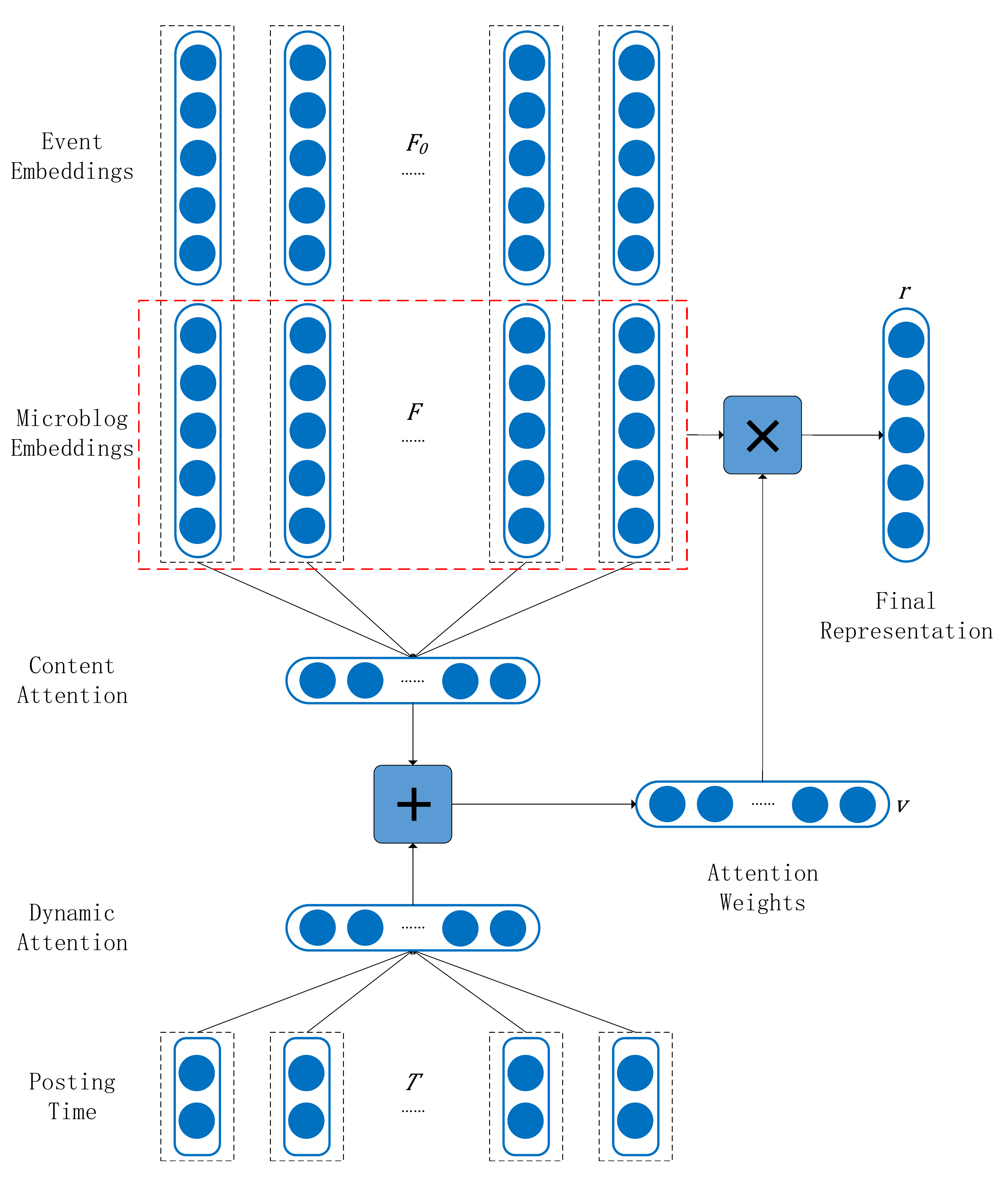}
\caption{Illustration of the proposed AIM model. The attention mechanism in AIM contains two parts: content attention and dynamic attention. Content attention is calculated based textual features of each microblog. Dynamic attention is related to the time interval between the posting time of a microblog and the beginning of the event.}
\label{fig:model}
\end{figure}

\subsection{Attention Mechanism for Misinformation Identification}

As we know, information on social media is usually full of noisy and outliers, which should be alleviated. Moreover, most of microblogs about an event have little contribution to the identification of misinformation. And useful information can be easily overwhelmed by useless information. Thus, a reliable misinformation identification method can benefit from several most significant microblogs. Meanwhile, the attention mechanism \cite{itti1998model} is suitable for selecting most significant components of information. Via the attention mechanism, components which contribute more to a specific task have large weights for satisfying the objective as much as possible. Thus, we propose a reliable automatic detection method to mine signification microblogs for identifying misinformation based on the attention mechanism.

For an event $e_i$ which contains $M^{e_i }  = \left[ {m_1^{e_i } ,m_2^{e_i } ,...,m_{n_{e_i}}^{e_i }} \right]$, we have a function ${\mathop{\rm A}\nolimits} \left( {{e_i}} \right)$ to calculate attention values for each microblog. To capture content features and dynamic properties simultaneously, we have content attention and dynamic attention which can be calculated via ${\mathop{\rm A_c}\nolimits} \left( {{e_i}} \right)$ and ${\mathop{\rm A_t}\nolimits} \left( {{e_i}} \right)$ respectively. And we will discuss them further in Section \ref{CA} and \ref{DA}. This process can be formulated as:
\begin{equation} \label{function}
{\mathop{\rm A}\nolimits} \left( {{e_i}} \right)  = {\mathop{\rm A_c}\nolimits} \left( {{e_i}} \right) + {\mathop{\rm A_t}\nolimits} \left( {{e_i}} \right) ~,
\end{equation}
which outputs a ${n_{e_i}}$-dimensional attention value vector. This vector can be further normalized, and thus we can obtain an attention weight vector:
\begin{equation} \label{weight}
\mathbf{v}^{e_i} = \mathop{\rm softmax}\nolimits \left( {\mathop{\rm A}\nolimits} \left( {{e_i}} \right) \right) ~,
\end{equation}
where $\mathbf{v}^{e_i} \in \mathbb{R}^{n_{e_i}}$, denoting weights for each microblog belonging to event $e_i$. Large attention weight indicates the corresponding microblog has significant effect on misinformation identification.

Then, weighted sum of all microblogs can be performed to generate the representation of the whole even:
\begin{equation} \label{weight sum}
\mathbf{r}^{e_i} = \mathbf{F}^{e_i } \mathbf{v}^{e_i} ~,
\end{equation}
where $\mathbf{F}^{e_i} \in \mathbb{R}^{d \times n_{e_i}}$ and $\mathbf{r}^{e_i} \in \mathbb{R}^{d}$. $\mathbf{r}^{e_i}$ denotes the final representation of event $e_i$. And $\mathbf{F}^{e_i }  = \left[ {\mathbf{f}_1^{e_i } ,\mathbf{f}_2^{e_i } ,...,\mathbf{f}_{n_{e_i}}^{e_i }} \right]$ denotes content features of micrblogs $M^{e_i }$. Here, we use textual embeddings of micrblogs as content features. And we apply para2vec \cite{le2014distributed} for extracting embeddings from micrblogs. Para2vec is an extended version of word2vec \cite{mikolov2013distributed}, and is a state-of-the-art method for extracting sentence embeddings. In this work, we empirically set the dimensionality of embeddings as $d=50$.

Finally, the prediction on $e_i$ can be made with a logistic regression:
\begin{equation} \label{prediction}
\hat l_{e_i } = \mathop{\rm sigmoid}\nolimits \left( \mathbf{W}^T \mathbf{r}^{e_i} + \mathbf{b} \right) ~,
\end{equation}
where $\mathbf{W} \in \mathbb{R}^{d}$ and $b \in \mathbb{R}$. $\hat l_{e_i} = 1$ means event $e_i $ is predicted to be misinformation and $\hat l_{e_i }  = 0$ otherwise. The larger the predicted value, the lower the credibility of event $e_i$.

\subsection{Content Attention} \label{CA}

Content features of microblogs are the most important factor for describing an event. These features can tell us what has happened and how people react. So, it is vital to generate attention values based on textual embeddings of microblogs $\mathbf{F}^{e_i }  = \left[ {\mathbf{f}_1^{e_i } ,\mathbf{f}_2^{e_i } ,...,\mathbf{f}_{n_{e_i}}^{e_i }} \right]$. Moreover, when people repost a microblog or make comments about a microblog, the original content is usually not included. A comment's correlation with the original content is also important for deciding its significance. So, we concatenate the textual embedding of the very first microblog, and obtain the embeddings of the event $\mathbf{F}_0^{e_i }  = \left[ {\mathbf{f}_1^{e_i } ,\mathbf{f}_1^{e_i } ,...,\mathbf{f}_{1}^{e_i }} \right]$. Based on microblog embeddings and event embeddings, we can calculate content attention value for microblogs.

First, we transfer above embeddings to a hidden space:
\begin{equation} \label{hidden}
{\mathbf{H}^{{e_i}}} = \tanh \left( {{\mathbf{W}_h}\left[ \begin{array}{l}
\mathbf{F}_0^{{e_i}}\\
{\mathbf{F}^{{e_i}}}
\end{array} \right]} \right) ~,
\end{equation}
where $\mathbf{W}^{h} \in \mathbb{R}^{d_h \times 2d}$ and $\mathbf{H}^{e_i} \in \mathbb{R}^{d_h \times n_{e_i}}$. $\mathbf{H}^{e_i}$ denotes the hidden representations of microblogs $M^{e_i } $. This hidden space allows the interaction between microblog embeddings and event embeddings, and leads a new space for calculating content attention values.

Then, we can calculate content attention values as:
\begin{equation} \label{content}
{\mathop{\rm A_c}\nolimits} \left( {{e_i}} \right)  = \mathbf{W}^T_a {\mathbf{H}^{{e_i}}} ~,
\end{equation}
where $\mathbf{W}^{a} \in \mathbb{R}^{d_h}$. These generated content attention values can be further managed in Equation \ref{function}.

\subsection{Dynamic Attention} \label{DA}

Besides content features, posting time of a microblog is also vital for deciding its significance. So, we incorporate dynamic attention, which can be determined related to the time interval between the posting time of a microblog and the beginning of the event. The beginning of an event is the posting time of the very first microblog of the corresponding event. For example, the beginning time of event $e_i$ is $t_1^{e_i}$. Accordingly, dynamic attention values can be calculated as:
\begin{equation} \label{dynamic}
{\mathop{\rm A_c}\nolimits} \left( {{e_i}} \right)  = \left[ {\mathbf{c}_{t_1^{e_i} - t_1^{e_i} } ,\mathbf{c}_{t_2^{e_i} - t_1^{e_i} } ,...,\mathbf{c}_{t_{n_{e_i}}^{e_i} - t_1^{e_i} }} \right] ~,
\end{equation}
where $\mathbf{c}_{t_j^{e_i} - t_1^{e_i} }$ denotes the dynamic attention value of the corresponding time interval ${t_j^{e_i} - t_1^{e_i} }$. These generated dynamic attention values can be further managed in Equation \ref{function}.

Furthermore, if we learn a distinct attention value for every possible continuous time interval value, we have to estimate a great number of parameters and the model tends to overfit. Here, following the method in \cite{liu2016strnn,liu2017multi,liu2016context}, we equally partition the range of all the possible time interval values into discrete bins. Specifically, in this work, the range of all the possible time interval values is partitioned into one-hour bins. Only the attention values of the upper and lower bounds of time bins are needed to be estimated in our model. For time interval values in a time bin, their attention values can be calculated via a linear interpolation.

Suppose we have an arbitrary time interval value $t_d = {t_j^{e_i} - t_1^{e_i} }$. Mathematically, the corresponding dynamic attention value $\mathbf{c}_{t_d }$ can be calculated as:
\begin{equation} \label{linear}
\mathbf{c}_{t_d}  = \frac{{\left[ {\mathbf{c}_{L(t_d)} (U(t_d) - t_d) + \mathbf{c}_{U(t_d)} (t_d - L(t_d))} \right]}}{{\left[ {(U(t_d) - t_d) + (t_d - L(t_d))} \right]}} ~,
\end{equation}
where $U(t_d)$ and $L(t_d)$ denote the upper bound and lower bound of time interval value $t_d$, $\mathbf{c}_{U(t_d)}$ and $\mathbf{c}_{L(t_d)}$ denote the dynamic attention values for $U(t_d)$ and $L(t_d)$ respectively. Such a linear interpolation method can solve the problem of learning attention values for continuous time intervals. To be noted, although the change of dynamic attention values in each discrete time bin is linear, the global change in the entire range of all the possible time interval values is nonlinear.

\subsection{Parameter Learning}

The proposed AIM model can be trained in an end-to-end way by backpropagation. The goal of training is to minimize the following error between $l_{e_i }$ and $\hat l_{e_i }$ for each event $e_i$:
\begin{equation} \label{loss}
J =  - \sum\limits_{i = 1}^n {{l_{{e_i}}}\ln {{\hat l}_{{e_i}}}}  - \sum\limits_{i = 1}^n {\left( {1 - {l_{{e_i}}}} \right)\ln \left( {1 - {{\hat l}_{{e_i}}}} \right)}  + \frac{\lambda }{2}\left\| {\mathbf{\theta }} \right\| ~,
\end{equation}
where $\lambda$ is the L2-regularization term, and ${\mathbf{\theta }} = \left\{ {\mathbf{W},\mathbf{b},{\mathbf{W}_h},{\mathbf{W}_a},\mathbf{c}} \right\}$ denoting all the parameters needed to be learned in AIM. Then, the derivations of J with respect to all the parameters can be calculated, and we can employ Stochastic Gradient Descent (SGD) to estimate the model parameters. The training procedure consists of two parts: the training of the attention mechanism which learns ${\mathbf{W}_h}$, ${\mathbf{W}_a}$ and $\mathbf{c}$, and the training of the logistic regression which learns $\mathbf{W}$ and $\mathbf{b}$. These two parts of training are done alternately. This process is repeated iteratively until the convergence is achieved.

\section{Experiments}

In this section, we first present our experimental settings. Then, we report experiment results of AIM on misinformation identification comparing with several state-of-the-art methods. We also investigate the impact of hyper-parameters in AIM. Moreover, we study the performance comparison on misinformation early detection.

\begin{table}[!tb]
  \centering
  \caption{Detailed statistics of the Weibo dataset and the Twitter dataset.}
    \begin{tabular}{lrr}
    \toprule
    statistics & \multicolumn{1}{l}{Weibo} & \multicolumn{1}{l}{Twitter} \\
    \midrule
    \#users & 2,746,818 & 491,229 \\
    \#microblogs & 3,805,656 & 1,101,985 \\
    \#events & 4,664 & 992 \\
    \#misinformation & 2,313 & 498 \\
    \#true information & 2,351 & 494 \\
    \bottomrule
    \end{tabular}%
  \label{data}%
\end{table}%

\subsection{Experimental Settings}

To evaluate the performance of AIM, following some representative previous works \cite{ma2015detect,ma2016detecting}, we conduct experiments on the \textbf{Weibo} dataset and the \textbf{Twitter} dataset. Detailed statistics of the two datasets can be found in Table \ref{data}. Misinformation in the Weibo dataset is collected from Weibo misinformation management center \footnote{http://service.account.weibo.com/?type=5\&status=4}, which reports various misinformation. And similar number of true information is collected by crawling microblogs of general threads that are not reported as misinformation. Misinformation and true information are confirmed on Snopes \footnote{http://www.snopes.com}, which is an online misinformation debunking service.

In our experiments, we randomly choose 10\% of events in each dataset for model tuning, and the rest 90\% are randomly assigned in a 3:1 ratio for training and testing. Empirically, the L2-regularization term is set to be $\lambda = 0.001$, the learning rate of SGD is set to be $0.01$, and the dimensionality of microblog embeddings is set to be $d=50$. And the dimensionality $d_h$ of hidden space for generating content attention is turned in our experiments.

Moreover, we adopt several evaluation metrics for evaluating the performance of AIM and other compared methods: \textbf{accuracy}, \textbf{precision}, \textbf{recall}, and \textbf{f1-score}. Accuracy is a standard metric for classification tasks, which is evaluated by the percentage of correctly predicted misinformation and true information. Precision, recall and f1-score are widely-used metrics for classification tasks, which are computed according to where correctly predicted misinformation or true information appear in the predicted list. The larger the values of the above evaluation metrics, the better the performance.

To demonstrate the effectiveness of AIM, several state-of-the-art methods are compared in our experiments:
\begin{itemize}
\item \textbf{GRU} has been incorporated for misinformation identification. A model based on two GRU hidden layers \cite{chung2015gated} and textual features in dynamic time windows achieves satisfactory performance in both misinformation identification and misinformation early detection \cite{ma2016detecting}.

\item \textbf{SVM-TS} is a linear Support Vector Machine (SVM) \cite{chang2011libsvm} classifier that uses time-series structures to model the variation of social context features \cite{ma2015detect}. Handcrafted features based on contents, users and propagation patterns are replicated.

\item \textbf{DT-Rank} is a ranking model based on decision tree \cite{quinlan1987simplifying} to identify trending misinformation \cite{zhao2015enquiring}. DT-Rank searches for enquiry phrases and cluster disputed factual claims, and ranks the clustered results based on statistical features.

\item \textbf{DTC} is a decision tree classifier \cite{castillo2011information}. It uses hand-crafted features based on the overall statistics of the posts, rather than temporal information.

\item \textbf{SVM-RBF} is a SVM-based model with the Radical Basis Function (RBF) kernel \cite{yang2012automatic}. It also uses hand-crafted features based on the overall statistics of the posts.

\item \textbf{DFC} is a random forest classifier \cite{ho1995random} using three parameters to fit the temporal posting volume curve \cite{kwon2013prominent}. Same handcrafted features are used as in SVM-TS.
\end{itemize}
For our proposed AIM model, we implement it with Python \footnote{https://www.python.org/} and Theano \footnote{http://deeplearning.net/software/theano/}. Moreover, versions of AIM without event embeddings and dynamic attention are also implemented and compared, for evaluating the compact of different components of AIM.

\subsection{Performance Comparison on Misinformation Identification}

Table \ref{tab:comparison} illustrates the performance comparison on misinformation comparison among AIM and several state-of-the-art models ont the Weibo and Twitter datasets. And the dimensionality of the hidden space for generating content attention is $d_h=40$. We can see that the performance ranking of misinformation identification methods is as follows: AIM, GRU, SVM-TS, RFC, DTC, SVM-RBF and DT-Rank. Compared with neural networks-based methods, i.e., AIM and GRU, the performance of other methods is relatively poor. These methods using handcrafted features or rules may not adapt to shape dynamic and underlying correlations in social media. In contrast, neural networks-based methods, AIM and GRU can learn high-level interactions among deep latent features, which can better model real-world scenarios.

Among those conventional methods, DT-Rank uses a set of regular expressions selected from signal microblog posts containing skeptical enquiries. But not all microblog posts in both Twitter and Weibo datasets involve these skeptical enquiries. These selected expressions are insufficient to conclude the information credibility. Moreover, SVM-TS and RFC incorporate the dynamic properties into conventional models, which helps outperform other compared methods like SVM-RBF and DTC. So, we can conclude that dynamic properties are important features for misinformation identification.

From the experimental results, we can clearly observe that GRU achieves the best performance among all the compared methods. And it is obvious that AIM obtains significant improvement over GRU. On the Weibo dataset, comparing with GRU, AIM improves the performance by $2.8\%$, $2.2\%$ and $3.4\%$ evaluated by accuracy, f1-score (misinformation) and f1-score (true information) respectively. On the Twitter dataset, the improvements become $3.9\%$, $2.3\%$ and $2.8\%$. Despite the fact that both AIM and GRU learn deep latent features from a sequence of groups of microblog posts, a trained GRU model possesses a constant recurrent transition matrix, which induces unchangeable propagations of sequence signals between every two consecutive time windows. However, in real-world scenarios, social media is so dynamic and complicated that the above constant recurrent transition matrix of the GRU model has its limitation to shape an adequate misinformation identification model. Moreover, the GRU model has a bias towards the latest elements that it takes as input \cite{mikolov2011extensions}. While key features of both misinformation and truth information do not necessarily appear at the rear part of an input sequence. Meanwhile, all the compared methods, including GRU, can not select significant microblogs for misinformation identification, and thus outliers and useless information will lower the performance. Via overcoming such a shortcoming, AIM shows its superiority in misinformation identification, which is proven by the experimental results.

\begin{table}[!tb]
\small
  \centering
  \caption{Performance Comparison on misinformation identification with $d_h = 40$ on the Weibo and Twitter datasets. \textbf{M} stands for misinformation, and \textbf{T} stands for true information. Results are evaluated by accuracy, precision, recall, and f1-score.}
    \begin{tabular}{ccccccccccc}
\toprule
    \multirow{2}[0]{*}{method} & \multirow{2}[0]{*}{class} & \multicolumn{4}{c}{Weibo}     &       & \multicolumn{4}{c}{Twitter} \\
          &       & accuracy & precision & recall & f1-score &       & accuracy & precision & recall & f1-score \\
\midrule
    \multirow{2}[0]{*}{DT-Rank} & M     & \multirow{2}[0]{*}{0.732 } & 0.738  & 0.715  & 0.726  &       & \multirow{2}[0]{*}{0.681 } & 0.711  & 0.698  & 0.704  \\
          & T     &       & 0.726  & 0.749  & 0.737  &       &       & 0.647  & 0.662  & 0.655  \\
    \multirow{2}[0]{*}{SVM-RBF} & M     & \multirow{2}[0]{*}{0.818 } & 0.822  & 0.812  & 0.817  &       & \multirow{2}[0]{*}{0.715 } & 0.698  & 0.809  & 0.749  \\
          & T     &       & 0.815  & 0.824  & 0.819  &       &       & 0.741  & 0.610  & 0.669  \\
    \multirow{2}[0]{*}{DTC} & M     & \multirow{2}[0]{*}{0.831 } & 0.847  & 0.815  & 0.831  &       & \multirow{2}[0]{*}{0.718 } & 0.721  & 0.711  & 0.716  \\
          & T     &       & 0.815  & 0.847  & 0.830  &       &       & 0.715  & 0.725 & 0.720  \\
    \multirow{2}[0]{*}{RFC} & M     & \multirow{2}[0]{*}{0.849 } & 0.786  & 0.959  & 0.864  &       & \multirow{2}[0]{*}{0.728 } & 0.742  & 0.737  & 0.740  \\
          & T     &       & 0.947  & 0.739  & 0.830  &       &       & 0.713  & 0.718  & 0.716  \\
    \multirow{2}[0]{*}{SVM-TS} & M     & \multirow{2}[0]{*}{0.857 } & 0.839  & 0.885  & 0.861  &       & \multirow{2}[0]{*}{0.745 } & 0.707  & \textbf{0.864} & 0.778  \\
          & T     &       & 0.878  & 0.830  & 0.857  &       &       & 0.809  & 0.618  & 0.701  \\
    \multirow{2}[1]{*}{GRU} & M     & \multirow{2}[1]{*}{0.908 } & 0.874  & \textbf{0.954}  & 0.912  &       & \multirow{2}[1]{*}{0.757 } & 0.732  & 0.815  & 0.771  \\
          & T     &       & \textbf{0.950}  & 0.862  & 0.904  &       &       & 0.788  & 0.698  & 0.771 \\
    \midrule
    AIM   & M     & \multirow{2}[0]{*}{0.934} & 0.920 & 0.943 & 0.931 &       & \multirow{2}[0]{*}{0.791} & 0.737 & 0.835 & 0.783 \\
    (-$\mathbf{F}_0$) & T     &       & 0.947 & 0.926 & 0.936 &       &       & 0.841 & 0.732 & 0.783 \\
    AIM   & M     & \multirow{2}[0]{*}{0.927} & 0.915 & 0.934 & 0.925 &       & \multirow{2}[0]{*}{0.770} & 0.725 & 0.826 & 0.772 \\
    (-${\mathop{\rm A_t}\nolimits}$) & T     &       & 0.939 & 0.922 & 0.930 &       &       & 0.826 & 0.725 & 0.772 \\
    \multirow{2}[1]{*}{AIM} & M     & \multirow{2}[1]{*}{\textbf{0.936}} & \textbf{0.922} & 0.945 & \textbf{0.934} &       & \multirow{2}[1]{*}{\textbf{0.796}} & \textbf{0.746} & 0.846  & \textbf{0.794} \\
          & T     &       & 0.949 & \textbf{0.928} & \textbf{0.938} &       &       & \textbf{0.851} & \textbf{0.754}  & \textbf{0.799}  \\
    \bottomrule
\end{tabular}%
  \label{tab:comparison}%
\end{table}%

\subsection{Impact of Event Embeddings and Dynamic Attention}

To investigate the impact of event embeddings $\mathbf{F}_0$ in content attention on misinformation identification, we implement a version of AIM without event embeddings, denoted as AIM (-$\mathbf{F}_0$). The corresponding performance on on misinformation identification is shown is Table \ref{tab:comparison}. Comparing with AIM, AIM (-$\mathbf{F}_0$) slightly decrease the performance on both datasets. This indicates that, a comment's correlation with the original content indeed has contribution to deciding its significance. But such effect is not very significant.

Similarly, to investigate the impact of dynamic attention ${\mathop{\rm A_t}\nolimits}$, a version of AIM, i.e., AIM (-${\mathop{\rm A_t}\nolimits}$) is implemented and compared in Table \ref{tab:comparison}. Comparing with AIM, AIM (-${\mathop{\rm A_t}\nolimits}$) decreases the accuracy by $0.9\%$ and $2.6\%$ on the Weibo dataset and the Twitter dataset respectively. It is obvious that the decay brought by AIM (-${\mathop{\rm A_t}\nolimits}$) is relatively large. This indicates that, the posting time of a microblog is important for deciding its significance, and dynamic attention is vital for misinformation identification. Moreover, AIM (-${\mathop{\rm A_t}\nolimits}$) can still outperform GRU, which means content attention itself can beat the compared methods.

\begin{figure*}[!tb]
\centering
\subfigure[Performances on the Weibo dataset.]{
\begin{minipage}[b]{0.48\textwidth}
\includegraphics[width=1\textwidth]{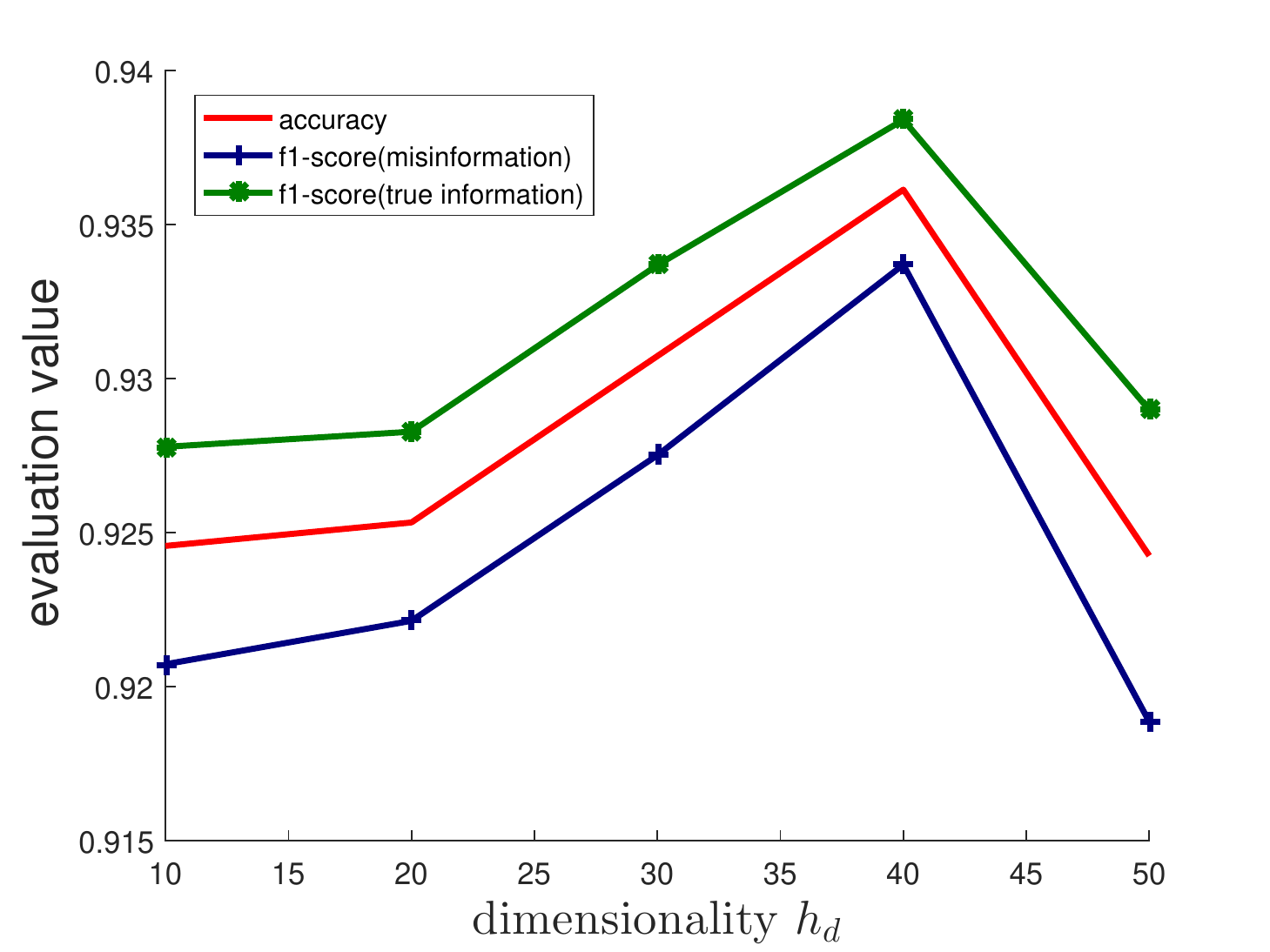}
\label{dim_weibo}
\end{minipage}
}
\hspace{-2mm}
\subfigure[Performances on the Twitter dataset.]{
\begin{minipage}[b]{0.48\textwidth}
\includegraphics[width=1\textwidth]{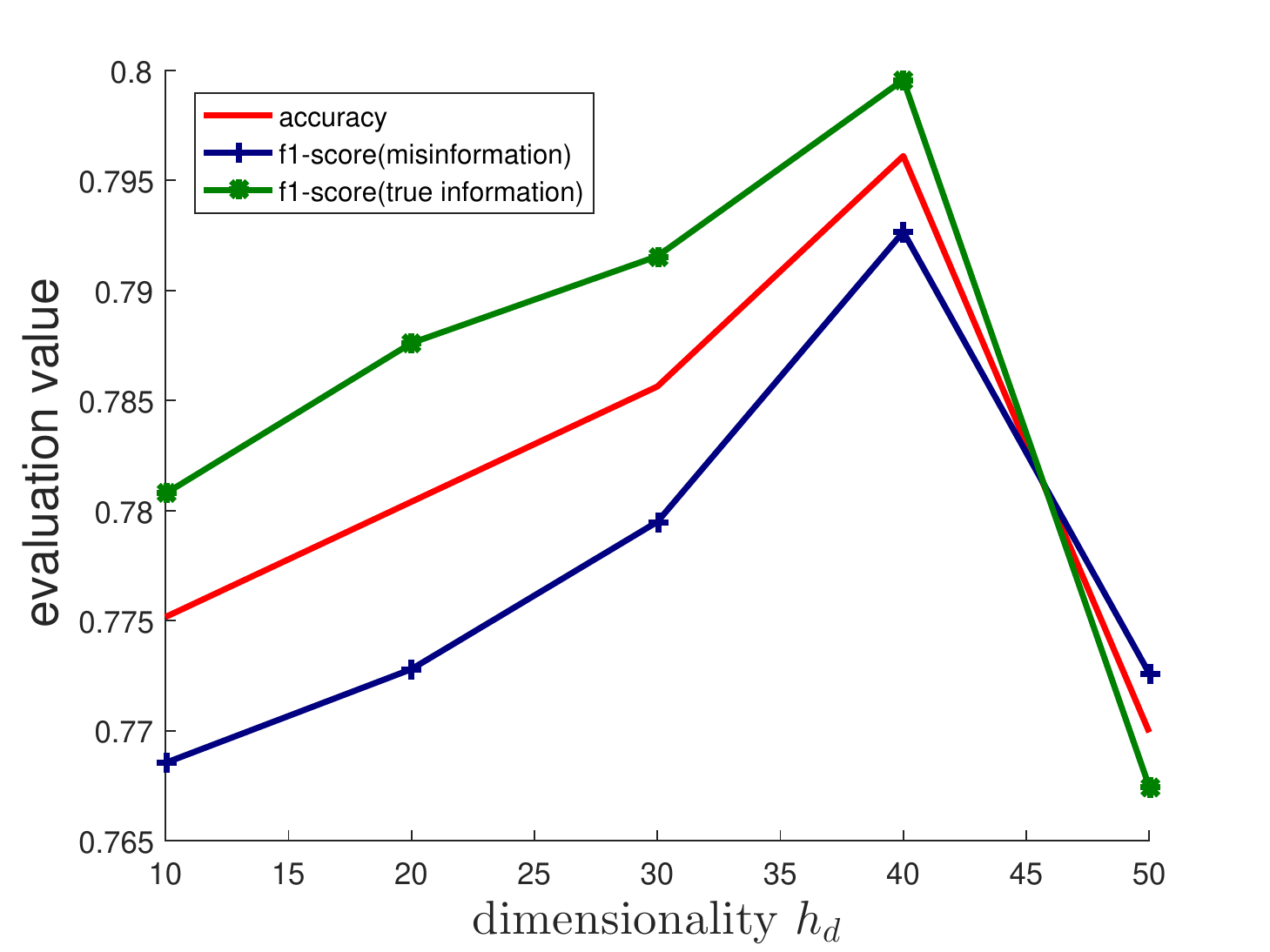}
\label{dim_twitter}
\end{minipage}
}
\caption{Performances of AIM on misinformation identification with varying dimensionality $d_h=[10,20,30,40,50]$. Results are evaluated by accuracy, and f1-score for both misinformation and true information.}
\label{fig:dim}
\end{figure*}

\subsection{Impact of Dimensionality}

The dimensionality $d_h$ of hidden space for generating content attention is a hyper-parameter in AIM. To investigate its impact on the performance of AIM, we illustrate the performance of AIM evaluated by accuracy and f1-score with varying $d_h=[10,20,30,40,50]$ in Figure \ref{fig:dim}. F1-score is evaluated for both misinformation and true information.

From the figure, on both datasets, we can clearly observe that, the performance of AIM increases rapidly from $d_h=10$. It achieves the best performance at $d_h=40$, and then decreases with the increasing dimensionality. Curves evaluated by accuracy, f1-score (misinformation) and f1-score (true information) share similar trends. From our observation, we select the best dimensionality of AIM as $d_h=40$, and report the corresponding results in the rest of our experiments. Moreover, these curves show that AIM is not very sensitive to the hidden dimensionality in a large range, and it can still outperforms the compared methods even not with the best dimensionality.

\begin{figure*}[!tb]
\centering
\subfigure[Performances on the Weibo dataset.]{
\begin{minipage}[b]{0.8\textwidth}
\includegraphics[width=1\textwidth]{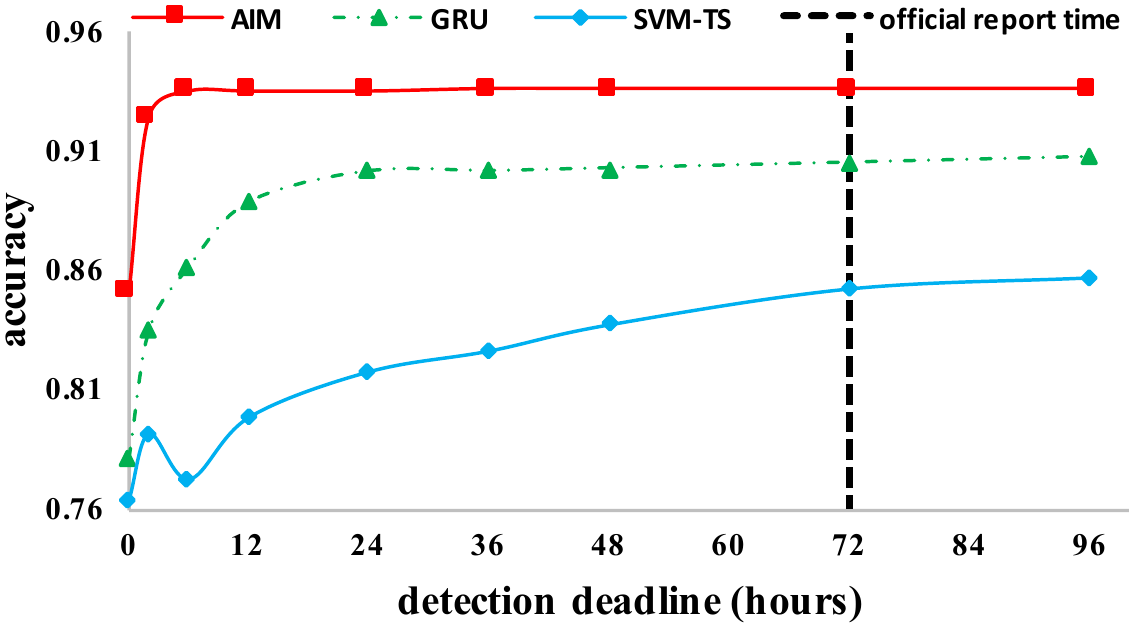}
\label{early_weibo}
\end{minipage}
}
\subfigure[Performances on the Twitter dataset.]{
\begin{minipage}[b]{0.8\textwidth}
\includegraphics[width=1\textwidth]{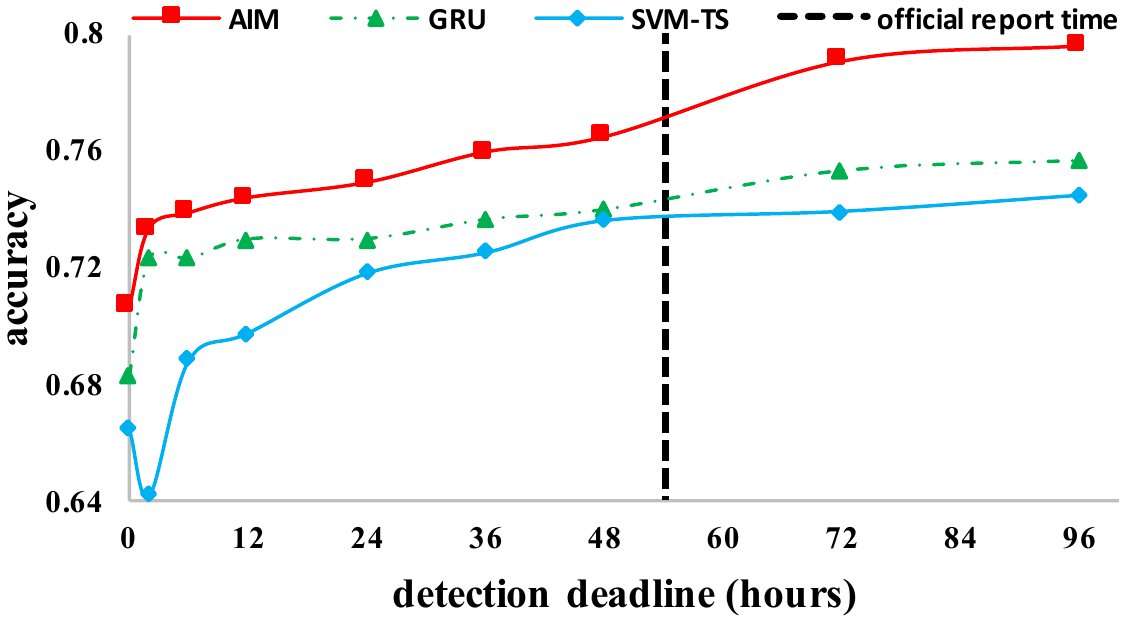}
\label{early_twitter}
\end{minipage}
}
\caption{Performances of AIM, GRU and SVM-TS on misinformation early detection with $d_h = 40$. Results are evaluated by accuracy. Official report time indicates the average time required for publicly reporting misinformation on the platform.}
\label{fig:early}
\end{figure*}

\subsection{Performance Comparison on Misinformation Early Detection}

Misinformation early detection is an important and practical task. We need to detect misinformation as early as possible. Thus, we can take immediate actions at the beginning stage of spreading of misinformation, and minimize the baneful influence. To investigate the performance of AIM on misinformation early detection, we select most competitive methods, i.e., GRU and SVM-TS with highest accuracies according to Table \ref{tab:comparison}, and illustrate their performance with varying detection deadlines in Figure \ref{fig:early}. The performance is evaluated by accuracy, and $d_h=40$. Moreover, conventional early detection tasks count on official announcements, which is the average reporting time over misinformation and announced by the debunking services such as Snopes and Sina community management center. So, we take official report time as a baseline.

As shown in the figure, the accuracy of most methods will experience a conspicuous climbing during the first few hours and then rise with different growth rates, convergence rates and convergence accuracies. For instance, accuracy curve of SVM-TS climbs slowly at early phase and gradually converge to a relatively low accuracy. Moreover, its accuracy curve still fluctuates after the official report time. While the accuracy curve of GRU climbs rapidly at early phase and converges to a much higher accuracy on a much earlier deadline than that of SVM-TS.

The proposed AIM models can reach relatively high accuracy at a very early time while other methods will take a longer time to achieve good performance. Furthermore, comparing with the performance of  GRU and SVM-TS, that of the proposed AIM model takes a relatively large lead at any phase. On the Weibo dataset, the accuracy of AIM can reach more than $90\%$, which is a very high accuracy and even larger than the performance of GRU on misinformation identification, in just two hours. At the official report time on the Weibo dataset and the Twitter dataset, the accuracy of AIM reaches about $93\%$ and $77\%$ respectively. These experimental results show that, the proposed AIM model is very practical for misinformation early detection.

Most state-of-the-art methods for early detection, such as GRU and SVM-TS, usually follow the intuitive paradigm to model time series features in sequences of microblog posts. But these time-series-based models are not qualified for practical early detection due to the conflict between the models and the task. Take GRU as an example. On one hand, the input sequence should be long enough to embody these possibly existing dynamic temporal signals to be captured by GRU. On the other hand, the practical early detection means limited input sequence can be used. The limited input sequence may not cover required dynamic temporal signals. So GRU may not be suitable for early detection of misinformation in some cases. With the attention mechanism, AIM identifies misinformation with several significant microblogs. This minimize the conflict between the models and the task, and make AIM a naturally suitable model for misinformation early detection.

\begin{figure*}[!tb]
\centering
\subfigure[Normalized dynamic attention values on the Weibo dataset.]{
\begin{minipage}[b]{0.48\textwidth}
\includegraphics[width=1\textwidth]{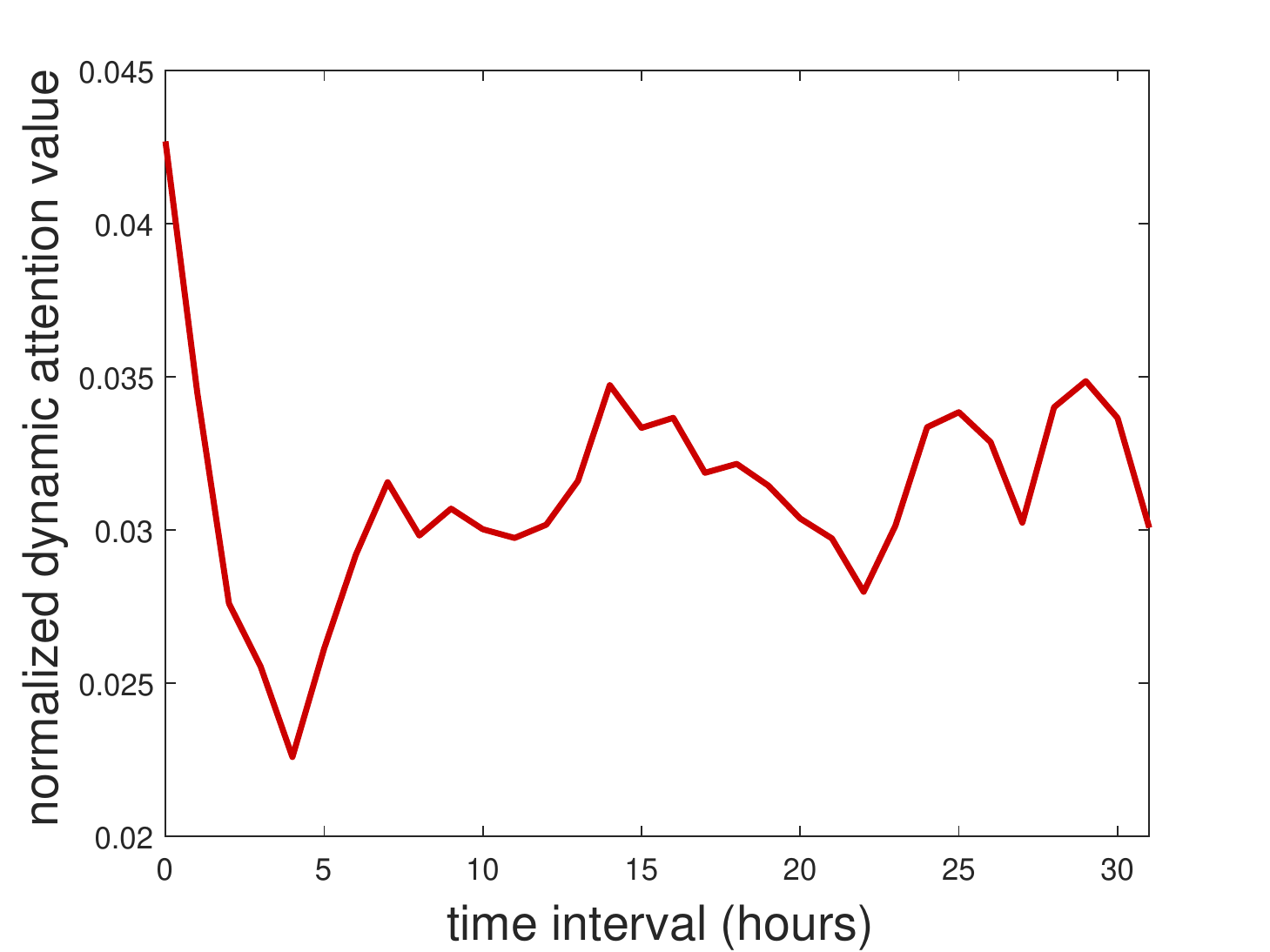}
\label{dynamic_weibo}
\end{minipage}
}
\hspace{-2mm}
\subfigure[Normalized dynamic attention values on the Twitter dataset.]{
\begin{minipage}[b]{0.48\textwidth}
\includegraphics[width=1\textwidth]{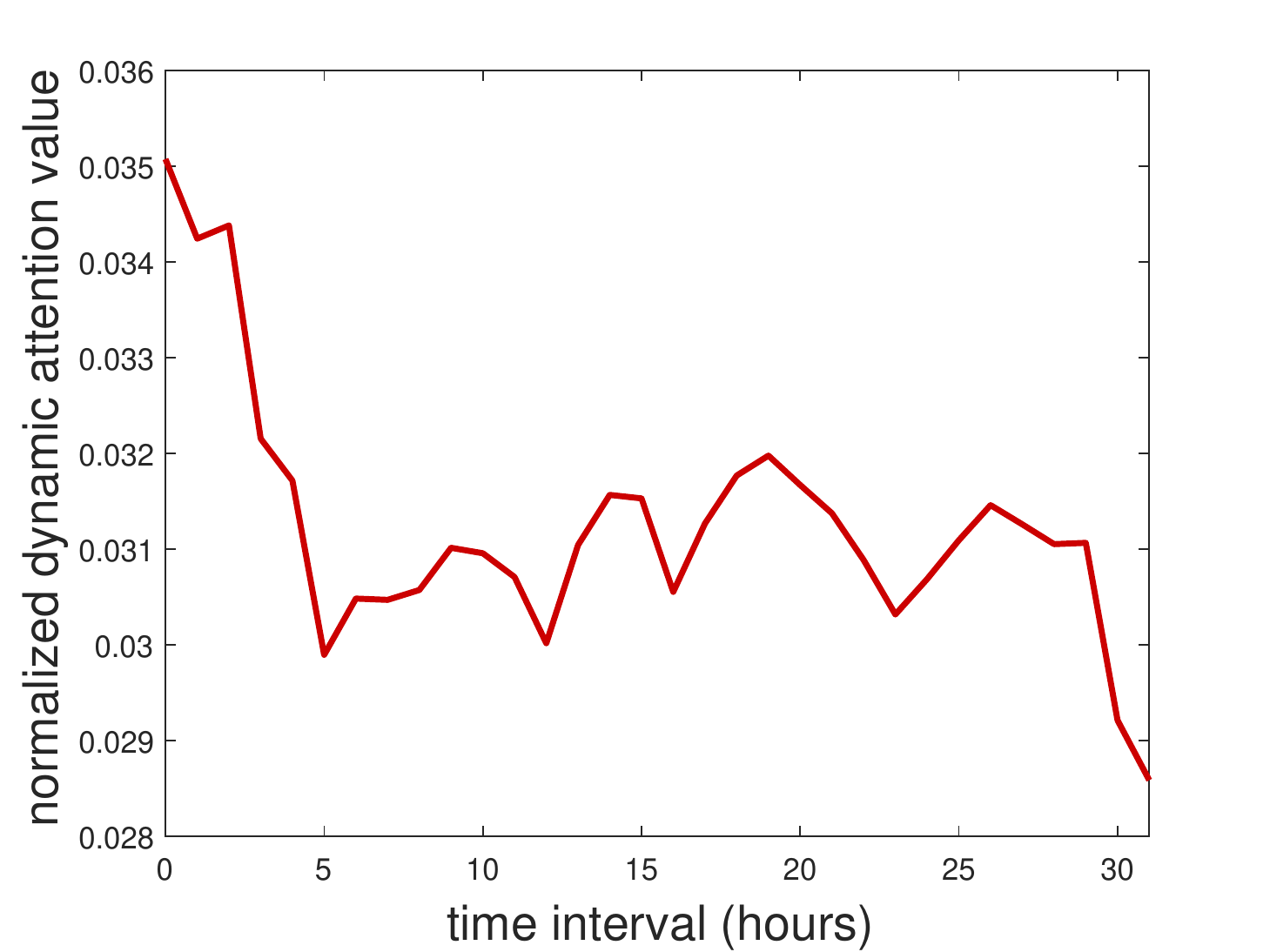}
\label{dynamic_twitter}
\end{minipage}
}
\caption{The illustration of the learned dynamic attention in AIM for different time intervals. The dynamic attention values are normalized. Larger the attention value, higher the significance.}
\label{fig:dynamic}
\end{figure*}

\section{Visualization}

In this section, we present the visualization of the leaned attention mechanism, for demonstrating the rationality of our proposed AIM model. First, we illustrate the curves of the learned dynamic attention values in AIM. Then, we pick several events, including both misinformation and true information, and illustrate microblogs with largest weights in each event.

\subsection{Visualization of Dynamic Attention}

In Figure \ref{fig:dynamic}, we illustrate the learned dynamic attention values in AIM on the Weibo and Twitter datasets. The dynamic attention values for different time intervals since the beginning of an event are normalized. Higher dynamic attention value indicates the corresponding time interval has larger significance for misinformation identification.

From the curves on the Weibo dataset and the Twitter dataset, we can draw similar conclusions. At the very beginning of an event, we have the largest dynamic attention values. This means that, the very first microblog, as well as some early comments, is very important for identify misinformation. Then, the attention values decreases rapidly till about 5 hours. This may indicate that, during this time period, people tend to repost and echo the message, and there is little useful information. Then, the curve starts to increase and achieve another high level. This shows that, at this moment, people starts to think and express their own attitudes, which may include suspicion, affirmation and denial. Finally, the curve tends to shock and then stay stable. Curves in Figure \ref{fig:dynamic} demonstrates the rationality of the learned dynamic attention values in AIM, and provides proofs for the contribution of dynamic attention on misinformation identification.

\subsection{Microblogs with Largest Weights}

\begin{table}[!tb]
  \centering\scriptsize
  \caption{Example 1 of misinformation in the Weibo dataset. Several microblogs with largest attention weights are illustrated.}
  {
    \begin{tabular}{cl}
    \toprule
    posting time & \multicolumn{1}{c}{content} \\
    \midrule
    \multirow{2}[0]{*}{2014/03/20 23:55} & Hearing from an Australian friend: The plane has been found in the international waters \\
          & near Perth. It is proven to be MH370 according to a major component of the plane. \\
    2014/03/22 01:25 & Is it reliable? \\
    2014/03/22 00:35 & Reposting unreliable information, what an expert! \\
    2014/03/22 00:49 & Hopefully it's not true. \\
    2014/03/22 09:48 & This can't be true. \\
    2014/03/22 00:04 & Waiting for official confirmation tomorrow. \\
    2014/03/22 07:33 & Really??? \\
    2014/03/22 00:17 & Let¡¯s watch the exact news tomorrow morning. Anyway, may God bless them! \\
    2012/08/22 00:05 & Is it true? \\
    \bottomrule
    \end{tabular}%
  }
  \label{tab:example_m1}%
\end{table}%

\begin{table}[!tb]
  \centering\scriptsize
  \caption{Example 2 of misinformation in the Weibo dataset. Several microblogs with largest attention weights are illustrated.}
  {
    \begin{tabular}{cl}
    \toprule
    posting time & \multicolumn{1}{c}{content} \\
    \midrule
    \multirow{2}[0]{*}{2012/08/29 19:20} & This mourning, 30 trucks full of coins arrived at Apple's headquarters. Samsung paid Apple \\
          & one-billion fine, with 20 billion coins! \\
    2012/08/29 19:22 & Is it true? \\
    2012/08/29 19:23 & Is it true... \\
    2012/08/29 19:36 & What? You must be kidding! I wonder how Steve Jobs feels. \\
    2012/08/29 22:38 & How can they get such a great amount of coins? \\
    2012/08/29 21:24 & Is this rumor or humor? Laughing... \\
    2012/08/30 01:54 & How cheating this is! \\
    \bottomrule
    \end{tabular}%
  }
  \label{tab:example_m2}%
\end{table}%

In Table \ref{tab:example_m1}-\ref{tab:example_t2}, we pick several events, containing both misinformation and true infroamtion, and illustrate several microblogs with largest attention weights belong to each event. Attention weights consist of both content attention and dynamic attention, as in Equation \ref{function}-\ref{weight}. From these examples, we can observe what kind of microblogs can contribute to misinformation identification the most.

\begin{table}[!tb]
  \centering\scriptsize
  \caption{Example 3 of misinformation in the Weibo dataset. Several microblogs with largest attention weights are illustrated.}
  {
    \begin{tabular}{cl}
    \toprule
    posting time & \multicolumn{1}{c}{content} \\
    \midrule
    2013/04/05 22:08 & China Mobile will charge Weixin and Weibo since September 1st. 10 Yuan per 500 messages. \\
    2013/04/05 22:09 & Surprising... Reliable news? \\
    2013/04/05 22:38 & Is it true? \\
    2013/04/05 22:12 & If this is true, I will not use Weixin and Weibo anymore. \\
    2013/04/06 08:24 & Luckily, I'm a user of China Telecommunications. \\
    2013/04/05 22:18 & Luckily, I'm not a user of China Mobile. \\
    2013/04/06 09:23 & @China Mobile Any confirmation? \\
    2013/04/06 16:33 & What kind of logic this is! \\
    2013/04/05 22:31 & Surprising! \\
    \bottomrule
    \end{tabular}%
  }
  \label{tab:example_m3}%
\end{table}%

\begin{table}[!tb]
  \centering\scriptsize
  \caption{Example 4 of misinformation in the Weibo dataset. Several microblogs with largest attention weights are illustrated.}
  {
    \begin{tabular}{cl}
    \toprule
    posting time & \multicolumn{1}{c}{content} \\
    \midrule
    \multirow{2}[0]{*}{2012/08/11 14:23} & At 8:20 this morning, a vicious explosion happened in Public Security Bureau of Jieshou City \\
          & Anhui Province. Seven policemen died on the spot. \\
    2012/08/11 17:09 & Is it true? \\
    2012/08/11 20:25 & Any confirmation? \\
    2012/08/11 15:30 & Official confirmation is need. \\
    2012/08/11 14:46 & Any official confirmation? \\
    2012/08/11 14:47 & No picture, no truth! \\
    2012/08/11 14:48 & Friends in Anhui, explain the real situation. \\
    2012/08/11 18:42 & Really??? \\
    \bottomrule
    \end{tabular}%
  }
  \label{tab:example_m4}%
\end{table}%

Table \ref{tab:example_m1}-\ref{tab:example_m4} illustrate four examples of misinformation. Example 1 is a piece of misinformation about MH370, saying it has crashed near Australia. Example 2 is an absurd joke about Samsung and Apple. Example 3 is a fake new policy of China Mobile. Example 4 talks about a fabricated explosion accidence. From these examples, it is clear that, the original microblog of an event has very large attention weight. This confirms to the common sense, because the original microblog details what has happened, which is important for understanding the event. Among other microblogs with large attention weights, most are about denying the event, questioning the real situation, suspecting the event or sarcasm. Obviously, the attention mechanism in AIM can mine microblogs that are most significant for misinformation. This make AIM a reliable misinformation detection approach, which can alleviated outliers and useless information.

Table \ref{tab:example_t1}-\ref{tab:example_t2} illustrate two examples of true information. Example 1 is about the birthday the panda Panpan. Example 2 is an incredible and funny story about a stupid thief. Though example 2 is a little absurd and hard to believe, it is a piece of true information and has been correctly classified by AIM. As in table \ref{tab:example_m1}-\ref{tab:example_m4}, the original microblog of true information has very large attention weight. But other significant microblogs are different from those in table \ref{tab:example_m1}-\ref{tab:example_m4}. Comments in table \ref{tab:example_t1}-\ref{tab:example_t2} are mostly talking about the even or the news itself. Based on significant microblogs, difference between misinformation and true information can be easily distinguished by AIM.

\section{Conclusions and Future Work}

In this paper, we propose a novel attention-based approach for misinformation identification on social media. The attention mechanism in our proposed AIM model contains two parts: content attention and dynamic attention. Content attention is calculated based textual features of each microblog. Dynamic attention is related to the time interval between the posting time of a microblog and the beginning of the event. Via aggregation of the content attention and the dynamic attention, we can obtain the final attention weights for each microblog belonging to an event. Weighted sum of these microblogs can be performed to generate the final representation of the whole event. The experimental results on two real datasets, i.e., the Weibo dataset and the Twitter dataset, show that AIM can outperform the state-of-the-art methods. And visualization of the leaned attention mechanism in AIM illustrates the rationality of our proposed model.

\begin{table}[!tb]
  \centering\scriptsize
  \caption{Example 1 of true information in the Weibo dataset. Several microblogs with largest attention weights are illustrated.}
  {
    \begin{tabular}{cl}
    \toprule
    posting time & \multicolumn{1}{c}{content} \\
    \midrule
    2015/11/28 16:27 & Panda Panpan's "100"-year-old birthday! Would you like to send her a heartfelt blessing? \\
    2015/11/28 16:30 & Long live! \\
    2015/11/28 16:49 & Surprising! Happy birthday, Panpan. \\
    2015/11/28 21:45 & Happy birthday. Smiling... \\
    2015/11/28 18:48 & She's pretty! \\
    2015/11/28 16:29 & Happy birthday, Panpan. \\
    2015/11/28 23:37 & When I was little, I have worn a skirt with Panpan on it. \\
    2015/11/28 17:50 & Memory in childhood. Happy birthday. \\
    \bottomrule
    \end{tabular}%
  }
  \label{tab:example_t1}%
\end{table}%

\begin{table}[!tb]
  \centering\scriptsize
  \caption{Example 2 of true information in the Weibo dataset. Several microblogs with largest attention weights are illustrated.}
  {
    \begin{tabular}{cl}
    \toprule
    posting time & \multicolumn{1}{c}{content} \\
    \midrule
    \multirow{2}[0]{*}{2015/11/18 08:50} & A thief was caught in a net bar, playing an online game. He asked the policemen to wait \\
          & for a second, and said he couldn't implicate his teammates. \\
    2015/11/18 09:28 & Great teammate! HaHaHa... \\
    2015/11/18 11:42 & Great teammate! \\
    2015/11/18 08:56 & Laughing... \\
    2015/11/18 19:30 & The facial expression is so funny. \\
    2015/11/18 09:18 & If I were his teammate, I would be touched. \\
    2015/11/18 09:19 & Your teammate is so lucky! \\
    2015/11/18 16:35 & Policeman: I'll play for you. You can follow my colleagues. \\
    \bottomrule
    \end{tabular}%
  }
  \label{tab:example_t2}%
\end{table}%

In the future, we plan to investigate the following directions. In AIM, multiple features, such as posted images and propagation structure, are not incorporated. So, we can incorporate more features in our model. Moreover, AIM is a static model, i.e., parameters in AIM do not change among different time periods. However, the trend of topics on social media is dynamic over time, and we usually have different hot topics in different time periods. Thus, it is necessary to incorporate time-aware or topic-aware mechanism in AIM.

\bibliographystyle{ACM-Reference-Format}
\bibliography{acm}

\end{document}